\documentclass[aip,showpacs,preprint]{revtex4}
\usepackage{graphics}
\begin{document}
\title{Inverse magnetocaloric effect in polycrystalline La$_{0.125}$Ca$_{0.875}$MnO$_{3}$}


\author{Anis Biswas}
\email{anis.biswas@saha.ac.in}
\author{Tapas Samanta}
\email{tapas.samanta@saha.ac.in}
\author{S. Banerjee}
\author{I. Das}
\email{indranil.das@saha.ac.in}
\affiliation{Saha Institute of Nuclear Physics,1/AF,Bidhannagar,
Kolkata 700 064, India}

\begin{abstract}
Recently the inverse magnetocaloric effect is observed for different compounds.
However there is very rare for any manifestation of the effect to be seen in manganites.
We have found inverse magnetocaloric effect in the case of polycrystalline 
La$_{0.125}$Ca$_{0.875}$MnO$_{3}$. Such phenomenon is attributed to
 the stabilization of antiferromagnetic state associated with inherent 
magnetic inhomogeneous phases for this compound.   
\end{abstract}
\maketitle
The study of the magnetocaloric effect (MCE) has been a subject of intense research owing to its possible application
 in magnetic refrigeration~\cite{pecharsky1,phan,pecharsky2,nature,tsapl1,
tsapl2,guo,tsjap,phanapl,tsjpcm,sun,tapas,abmce1,abmce2,phanjap,abmce3,abmce4,jap}. It has been widely observed that
 the
 magnetic entropy reduces  with the application of magnetic field  for different materials including some paramagnetic salts~\cite{pecharsky1}. The cooling can 
be possible by 
utilizing those materials subject to expose them in the magnetic field 
followed by demagnetizing
 them adiabatically. However, there is recent report on the discovery of materials with an inverse situation in which,
 the magnetic configuration entropy increases due to the application
 of the magnetic field~\cite{nature}. Such an effect is known as inverse magnetocaloric effect
 (IMCE). For the materials exhibiting IMCE, the attaining of low temperature can be possible by only just magnetizing them adiabatically~\cite{nature}. 
Some of the examples of such materials are NiMnSn, FeRh, TbNiAl$_{4}$, DySb,
 Tb$_{2}$Ni$_{2}$Sn, NiMnSb etc~\cite{nature,nimnsn,ferh,tbnial4,tbnial4a,dysb,
tbni2sn, nimnsb}.  
The materials which show IMCE, can be used as heat-sink for heat generated when
 a conventional magnetocaloric material is magnetized before cooling by demagnetization under adiabatic condition~\cite{nature,jap}.
 The refrigeration efficiency can be enhanced by using materials
 exhibiting IMCE in composites with conventional magnetic refrigerants~\cite{nature}.  
Therefore searching of suitable materials, which display IMCE is an important issue in 
the ongoing research related to the magnetic refrigeration.   
 Our present study is based on the magnetocaloric property of polycrystalline La$_{0.125}$Ca$_{0.875}$MnO$_{3}$. We have observed IMCE with quite
 large value of change of magnetic entropy (-$\Delta{S}$) in this compound.

The perovskite manganites with general formula R$_{1-x}$B$_{x}$MnO$_{3}$ (R is 
rare-earth, B is bivalent ion) are considered potential magnetic refrigerants
~\cite{phan,abmce1,abmce2,phanjap,abmce3,abmce4}. La$_{1-x}$Ca$_{x}$MnO$_{3}$ is a
 manganite system, which has a very rich phase diagram depending
 on the values of x~\cite{tokura}. Although some works 
 regarding the magnetocaloric properties of this system with 
x$\sim$ $0.2-0.5$ are reported, there is hardly any such study 
for high doping 
concentration of bivalent ion~\cite{phan}. In fact,
 a little attention has been paid to the study of MCE for other manganite 
systems in high doping region (i.e, high value of x) also. 
We have chosen 
La$_{0.125}$Ca$_{0.875}$MnO$_{3}$ for two
 main reasons. Firstly, it is a system with high doping concentration
 of bivalent ion. Secondly, its position in the phase diagram of 
La$_{1-x}$Ca$_{x}$MnO$_{3}$ is at phase boundary between antiferromagnetic (AFM) and canted antiferromagnetic (CAF) phases~\cite{tokura}. 
Therefore this compound provides an opportunity to study the effect of the phase boundary on magnetocaloric property of a system as well.
There are reports of the observation of IMCE in the chrage ordered
 systems such as  Pr$_{0.5}$Sr$_{0.5}$MnO$_{3}$,
Nd$_{0.5}$Sr$_{0.5}$MnO$_{3}$~\cite{chenepl,sande}. For that systems, 
charge order transition and antiferromagnetic transition occurs simultaneously
~\cite{chenepl,sande}.
In complete contrast to those cases, in the present system charge ordering
 hardly occurs. In spite of this, the system exhibits IMCE.

The polycrystalline La$_{0.125}$Ca$_{0.875}$MnO$_{3}$ was prepared by sol-gel
 method. The details of the solgel method has been described in our previous
 article~\cite{ab1}. At the end of the sol-gel process, the
 decomposed gel was annealed at $1400^{o}$C for $36$ hours. The x-ray powder
 diffraction study has confirmed the formation of the sample with single
 crystallographic phase with PNMA space group. The lattice parameters
 are determined as, a = $5.347 \AA$, b = $7.442 \AA$, and c = $5.318 \AA$ . 

A commercial SQUID magnetometer was utilized for magnetization study. The
 temperature dependence of dc susceptibility (Fig. 1) shows clear 
antiferromagnetic
 transition at $\sim 120$ K. The transition temperature is consistent
 with the phase diagram of the sample~\cite{tokura}. The magnetization 
measurement has been performed in the presence of $100$ Oe magnetic field in zero
 field cooled protocol.
We have also done the specific heat study using semi-adiabatic heat pulse 
method.
The transition is manifested by observed maxima in the temperature 
dependence of specific heat as well (inset of Fig. 1).
\begin{figure} 
\resizebox{8.0cm}{!}
{\includegraphics{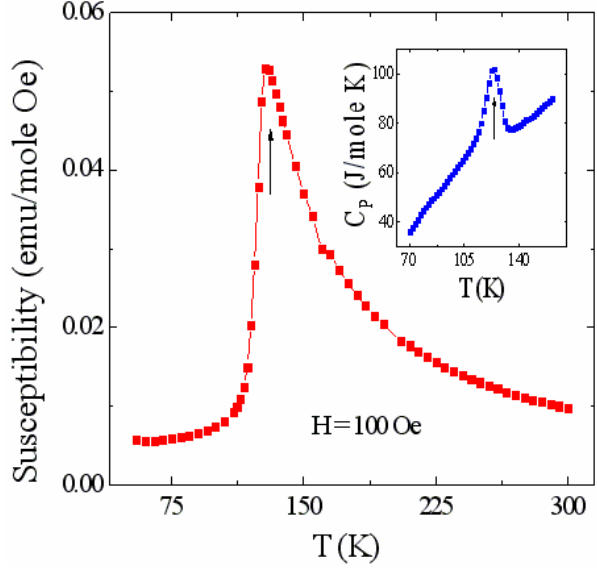}}
\caption{Temperature dependence of dc susceptibility for polycrystalline
 La$_{0.125}$Ca$_{0.875}$MnO$_{3}$ in the precence of $100$ Oe magnetic field.
Inset: temperature dependence of specific heat for this sample in the absence
 of magnetic field.
Antiferromagnetic transition is indicated by arrow.}
\end{figure}
\begin{figure}
\resizebox{8.0cm}{!}
{\includegraphics{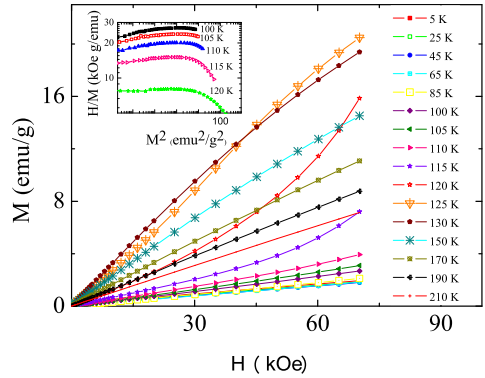}}
\caption{Magnetic field dependence of magnetization for polycrystalline
 La$_{0.125}$Ca$_{0.875}$MnO$_{3}$ at different temperatures.
Inset: H/M versus M$^{2}$ for the same sample at different temperatures around
 transition temperature.}
\end{figure}

The isothermal magnetic field dependence of magnetization [M(H)] at different
 temperatures has been studied for the sample (Fig. 2). To get intricate
 details of M(H), we have examined Banerjee's plot ~\cite{banerjee} i.e., 
H/M vs. M$^{2}$ behavior around transition temperature (inset, Fig. 2). The negative slope
 of Banerjee's plot is evident in high magnetic field region (above$\sim 35$ kOe), which is characteristic of first order transition~\cite{banerjee}.
From  the isothermal M(H) curves, the change of the magnetic entropy
(-$\Delta{S}$) was estimated for
various magnetic fields by using the Maxwell's relation~\cite{pecharsky1},
\begin{equation}
\left(\frac{\partial{S}}{\partial{H}}\right)_{T} = \left( \frac{\partial{M}}{\partial{T}}\right)_{H}
\end{equation}
 The temperature dependence of -$\Delta{S}$ for different magnetic field
 has been shown in Fig. 3. A minimum in -$\Delta{S}$(T) has been
 observed at $\sim 120$ K, where antiferromagnetic transition occurs.
One important feature of -$\Delta{S}$(T) is that the value of 
-$\Delta{S}$ remains negative for all the magnetic field at transition 
temperature for this sample. The negative value of -$\Delta{S}$ increases 
with
 the increase of magnetic field (i.e., the magnetic configuration entropy
 increases)  and it reaches  
$\sim -6.4$ J/kg K for the magnetic field change $0-70$ kOe. The increase
 of the negative value of -$\Delta{S}$  with magnetic field at transition 
temperature has been shown in inset (b) of Fig. 3. From the magnetocaloric
 behavior of the sample, it seems that although
 there is change of slope in Banerjee's plot at high magnetic field, the
 antiferromagnetism still exists in the field up to $\sim 70$ kOe. 
Recently, Ranke et al., put forward a theoretical framework of
 the magnetocaloric properties of antiferromagnetic system~\cite{ranke}.
The temperature dependence of -$\Delta{S}$ for 
La$_{0.125}$Ca$_{0.875}$MnO$_{3}$ follow that theoretical model quite
 convincingly. Previously, we have also observed small negative
 -$\Delta{S}$ at antiferromagnetic transition temperature for other manganite 
system at magnetic field below the required field for quenching the 
antiferromagnetism as well~\cite{abmce2}.      
\begin{figure}
\resizebox{10.0cm}{!}
{\includegraphics{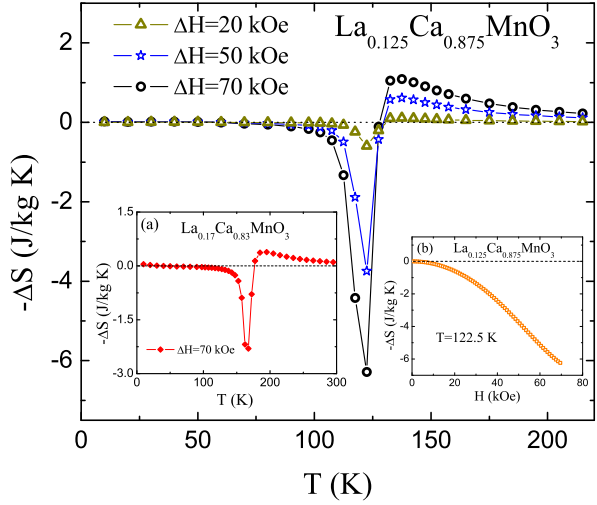}}
\caption{The temperature dependence of -$\Delta${S} for 
$\Delta${H} = $20$, $50$, and $70$ KOe in the case of La$_{0.125}$Ca$_{0.875}$MnO$_{3}$. Large change in  magnetic entropy has been observed around
 transition temperature. Inset (a): -$\Delta${S} vs. T for $\Delta$H = 70 kOe
 in the case of La$_{0.17}$Ca$_{0.83}$MnO$_{3}$
Inset (b): the magnetic field dependence of -$\Delta${S}
 around the transition temperature for La$_{0.125}$Ca$_{0.875}$MnO$_{3}$.}
\end{figure}

 Now question arises about the origin of
 the enhancement of magnetic configuration entropy with the increase of 
magnetic field.  
According to the phase diagram (temperature vs x) of 
La$_{1-x}$Ca$_{x}$MnO$_{3}$, La$_{0.125}$Ca$_{0.875}$MnO$_{3}$ situates
 at the phase boundary between antiferromagnetic and 
inhomogeneous canted antiferromagnetic state (CAF)~\cite{tokura}.
 It is obvious that CAF phase may have influence on the magnetic
 state of the sample. As
 a result of which,  magnetically inhomogeneous phase
 with mixed magnetic exchange interactions can be stabilized at the 
antiferromagnetic transition temperature. 
The enhancement of magnetic configuration entropy with the 
application of magnetic field can occur for such system giving
 rise to  IMCE~\cite{nature} with large value of -$\Delta{S}$.
We have also studied magnetocaloric property of another 
La$_{1-x}$Ca$_{x}$MnO$_{3}$, with slightly different value of x (x$\sim 0.83$).
That sample is of polycrystalline form and prepared in similar condition
 as La$_{0.125}$Ca$_{0.875}$MnO$_{3}$. IMCE is also observed for that compound
 at its antiferromagnetic transition temperature [inset (a), Fig. 3]. However 
the value of 
-$\Delta{S}$ is considerable less for La$_{0.17}$Ca$_{0.83}$MnO$_{3}$
 in comparison with La$_{0.125}$Ca$_{0.875}$MnO$_{3}$. According
 to the phase diagram of La$_{1-x}$Ca$_{x}$MnO$_{3}$, the magnetic transition
 in the case of La$_{0.17}$Ca$_{0.83}$MnO$_{3}$ is paramagnetic to 
antiferromagnetic, in which there is no influence of CAF phase~\cite{tokura}.
From the comparison of the magnetocaloric properties of the two compounds,
 it can be argued that the presence of magnetically inhomogeneous CAF state
 plays a vital role in IMCE and the magnetic entropy change becomes 
significantly enhanced as because of the influence of such state.

To summarize, we have observed IMCE in the case of polycrystalline 
La$_{0.125}$Ca$_{0.875}$MnO$_{3}$ with large value of -$\Delta{S}$ at 
antiferromagnetic transition temperature.
 Possibly,
 the stabilization of inhomogeneous magnetic state for this compound at its
antiferromagnetic transition temperature causes the increase in magnetic entropy
in the presence of magnetic field. The observation of the enhancement of
 magnetic configuration entropy with the increase of magnetic field is very rare
 especially for manganite systems.


\end{document}